\documentstyle[preprint,pra,floats,aps,psfig]{revtex}
\tightenlines

\begin{document} \draft
%----------------------------------------------------------------------

\title{\Large \bf Iwasawa Effects in Multi-layer Optics}

\author{Elena Georgieva\footnote{electronic address:
elena@physics.georgetown.edu}}
\address{Department of Physics, Georgetown University, Washington, DC
20057}

\author{Y. S. Kim\footnote{electronic address: yskim@physics.umd.edu}}
\address{Department of Physics, University of Maryland, College Park,
Maryland 20742}

\maketitle

\begin{abstract}
There are many two-by-two matrices in layer optics.  It is shown that
they can be formulated in terms of a three-parameter group whose
algebraic property is the same as the group of Lorentz transformations
in a space with two space-like and one time-like dimensions, or the
$Sp(2)$ group which is a standard theoretical tool in optics.  Among
the interesting mathematical properties of this group, the Iwasawa
decomposition drastically simplifies the matrix algebra under certain
conditions, and leads to a concise expression for the S-matrix for
transmitted and reflected rays.  It is shown that the Iwasawa effect
can be observed in multi-layer optics, and a sample calculation of the
S-matrix is given.

\end{abstract}


\vspace{5mm}

\narrowtext

\section{Introduction}\label{intro}

In a series of recent papers~\cite{hkn97josa,hkn97}, Han , Kim and
Noz have formulated polarization optics in terms
of the two-by-two and four-by-four representations of the six-parameter
Lorentz group.  They noted that the Lorentz group properties
can be found in optical materials.  Indeed, there are many two-by-two
matrices in layer optics~\cite{azzam77,monzon96,georg97}.  In this paper,
we reorganize them within the framework of the Lorentz group.  We
then derive a mathematical relation which can be tested experimentally.
If a light wave hits a flat surface, a part of this beam becomes
reflected and the remaining part becomes transmitted.

If there are multi-layers, this process repeats itself at each boundary.
There has been a systematic approach to this problem based on the
two-by-two S-matrix formalism~\cite{azzam77,monzon96,georg97}.  This
S-matrix consists of boundary and phase-shift matrices.  The phase-shift
matrices are complex and the S-matrix is in general complex.

However, in this paper, we show first these complex matrices can be
systematically transformed into a set of real traceless matrices with
three independent parameters.  Then we can use the well-established
mathematical procedure for them.  This procedure is called the $Sp(2)$
group whose algebraic property is the same as that of the $SU(1,1)$
group which occupies a prominent place in optics from squeezed states
of light~\cite{yuen76}.  However, the most pleasant aspect of the
$Sp(2)$ group is that its algebras consist only of two-by-two matrices
with real elements.  When applied to a two-dimensional plane, they
produce rotations and squeeze transformations~\cite{kim00job}.

It is known that these simple matrices produce some non-trivial
mathematical results, namely Wigner rotations and Iwasawa
decompositions~\cite{iwa49}.  The Wigner rotation means a rotation
resulting from a multiplication of three squeeze matrices, and the
Iwasawa decomposition means that a product of squeeze and rotation
matrices, under certain conditions, leads to a matrix with one vanishing
off-diagonal elements.  This leads to a substantial simplification
in mathematics and eventually leads to a more transparent comparison
of theory with experiments.  This decomposition has been discussed in
the literature in connection with polarization
optics~\cite{simon98,hkn99}.  In this paper, we study applications
of this mathematical device in layer optics.

There are papers in the literature on applications of the Lorentz
group in layer optics~\cite{hkn97,monzon99}, but these papers are
concerned with polarization optics.  In this paper, we are dealing
with reflections and transmissions of optical rays.  We show that
layers with alternate indexes of refraction can exhibit an Iwasawa
effect and provide a calculation of the transmission and reflection
coefficients.  It is remarkable that the Lorentz group can play as
the fundamental scientific language even in the physics of reflections
and transmissions.

In Sec.~\ref{formul}, we formulate the problem in terms of the
S-matrix method widely used in optics~\cite{azzam77}.
In Sec.~\ref{math}, this S-matrix formalism is translated into the
mathematical framework of the $Sp(2)$ group consisting of two-by-two
traceless matrices with real elements.   We
demonstrate that there is a subset of these matrices with one
vanishing non-diagonal element.  It is shown possible to produce
this set of matrices from multiplications of the matrices in the
original set.  This is called the Iwasawa decomposition.
In Sec. \ref{exp}, we transform the mathematical formalism of the
Iwasawa decomposition into the real world, and calculate the
reflection and transmission coefficients which can be measured in
optics laboratories.

\section{Formulation of the Problem}\label{formul}

Let us start with the S-matrix formalism of the layer optics.
If a beam is incident on a plane boundary of a medium with a different
index of refraction, the problem can be formulated in terms of two-by-two
matrices~\cite{azzam77,georg97}.  If we write the column vectors
\begin{equation}
\pmatrix{E_{1}^{(+)} \cr E_{1}^{(-)}} , \qquad
\pmatrix{E_{2}^{(+)} \cr E_{2}^{(-)}} ,
\end{equation}
for the incident, with superscript~(+), and reflected, with
superscript~(-), for the beams in the first and second media respectively,
then they are connected by the two-by-two S-matrix:
\begin{equation}\label{smat}
\pmatrix{E_{1}^{(+)} \cr E_{1}^{(-)}} =
\pmatrix{S_{11} & S_{12} \cr S_{21} & S_{22} }
\pmatrix{E_{2}^{(+)} \cr E_{2}^{(-)}} .
\end{equation}
Of course the elements of the above S-matrix depend on reflection and
transmission coefficients~\cite{azzam77}.

%------------------------------------------------------------------
\begin{figure}[thb]
\centerline{\psfig{figure=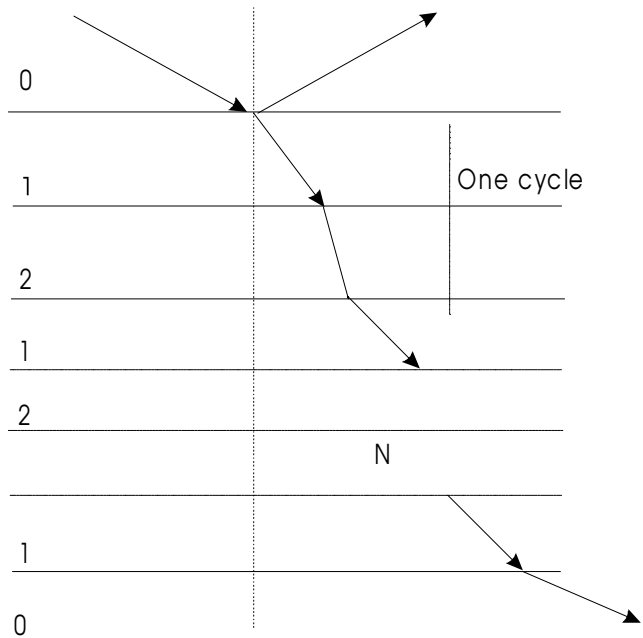,angle=0,height=100mm}}
\caption{Multi-layer system.  A light beam is incident on the first
boundary, with transmitted and reflected rays.  The transmitted ray
goes through the first medium and hits the second medium again with
reflected and transmitted rays.  The transmitted ray goes through
the second medium and hits the first medium.  This cycle continues
$N$ times.}\label{multi}

\end{figure}

%------------------------------------------------------------------

Let us consider a light wave incident on a flat surface, then it
is decomposed into transmitted and reflected rays.  If $E_{1}^{(+)}$
is the incident ray, the transmitted ray is $E_{2}^{(+)}$, with
\begin{equation}
E_{2}^{(+)} = t_{12} E_{1}^{(+)}, \qquad
E_{1}^{(-)} = r_{12} E_{1}^{(+)} .
\end{equation}
Thus, the S-matrix takes the form~\cite{azzam77}
\begin{equation}
\pmatrix{E_{1}^{(+)} \cr E_{1}^{(-)}} =
\pmatrix{1/t_{12} &  r_{12}/t_{12} \cr r_{12}/t_{12} & 1/t_{12}}
\pmatrix{E_{2}^{(+)} \cr 0} .
\end{equation}
If the ray comes from the second medium in the opposite direction, the
same matrix can be used for
\begin{equation}
\pmatrix{0 \cr E_{1}^{(-)}} =
\pmatrix{1/t_{12} & r_{12}/t_{12} \cr r_{12}/t_{12} & 1/t_{12}}
\pmatrix{E_{2}^{(+)} \cr E_{2}^{(-)}} .
\end{equation}
Since the magnitude of the reflection coefficient is smaller than one,
and since $t_{12}^{2} + r_{12}^{2} = 1$, we can write the above matrix
as
\begin{equation}\label{surf}
\pmatrix{\cosh\eta & \sinh\eta \cr \sinh\eta & \cosh\eta } ,
\end{equation}
with
\begin{equation}\label{coeff}
r_{12} = \tanh\eta , \qquad t_{12} = 1/\cosh\eta .
\end{equation}
Since this describes both the reflection and transmission at the
boundary, shall call this matrix ``boundary matrix''~\cite{monzon00}.
It is a uni-modular matrix (determinant = 1).  The mathematics
of this form is well known.  It can perform Lorentz boosts when
applied to the longitudinal and time-like coordinates.  Recently,
it has been observed that it performs squeeze transformations when
applied to the two-dimensional space of $x$ and $y$~\cite{kim00job}.

Next, if the ray travels within a given medium from one inner-surface
to the other surface~\cite{azzam77}
\begin{equation}\label{phase}
\pmatrix{E_{a}^{(+)} \cr E_{a}^{(-)}} =
\pmatrix{e^{-i\delta}  &  0 \cr 0 & e^{i\delta} }
\pmatrix{E_{b}^{(+)} \cr E_{b}^{(-)} } ,
\end{equation}
where the subscripts $a$ and $b$ are for the initial and final
surfaces respectively.  The above expression tells there is a phase
difference of $2\delta$ between the rays.  This phase difference
depends on the index of refraction, wavelength and the angle of
incidence~\cite{azzam77}.

In this paper, we consider a multi-layer system consisting of two
media with different indexes of refraction as is illustrated in
Fig. \ref{multi}.  Then, the system consists of many boundaries and
phase-shift matrices.  After multiplication of all those matrices,
the result will be one two-by-two matrix which we introduced as the
S-matrix in Eq.(\ref{smat}).  We are interested in this paper when
this matrix takes special forms which can be readily tested
experimentally.

If the ray hits the first medium from the air, as is illustrated
in Fig 1, we write the matrix as
\begin{equation}\label{s01}
\pmatrix{\cosh\lambda & \sinh\lambda \cr \sinh\lambda & \cosh\lambda} .
\end{equation}
Within the first medium, the phase shift matrix becomes
\begin{equation}\label{ps1}
\pmatrix{e^{-i\phi} & 0 \cr 0 & e^{i\phi}} .
\end{equation}
When the beam hits the surface of the second medium, the corresponding
matrix is
\begin{equation}\label{s12}
\pmatrix{\cosh\eta & \sinh\eta \cr \sinh\eta & \cosh\eta } .
\end{equation}
Within the second medium, we write the phase-shift matrix as
\begin{equation}\label{ps2}
\pmatrix{e^{-i\xi} & 0 \cr 0 & e^{i\xi}} .
\end{equation}
Then, when the beam hits the first medium from the second
\begin{equation}\label{s21}
\pmatrix{\cosh\eta & -\sinh\eta \cr -\sinh\eta & \cosh\eta } .
\end{equation}
But if the thickness of the first medium is zero, and the beam
exists to the air, then the system goes through the boundary matrix
\begin{equation}\label{s10}
\pmatrix{\cosh\lambda & -\sinh\lambda \cr -\sinh\lambda &
 \cosh\lambda } .
\end{equation}
The net result is
\begin{equation}
\pmatrix{\cosh\lambda & \sinh\lambda \cr \sinh\lambda &
 \cosh\lambda }
\pmatrix{\alpha & \beta \cr \gamma & \delta}
\pmatrix{\cosh\lambda & -\sinh\lambda \cr -\sinh\lambda &
 \cosh\lambda } ,
\end{equation}
with
\begin{eqnarray}
&{}&\pmatrix{\alpha & \beta \cr \gamma & \delta}
  = \pmatrix{e^{-i\phi} & 0 \cr 0 & e^{i\phi}} \pmatrix{\cosh\eta &
  \sinh\eta \cr \sinh\eta & \cosh\eta } \nonumber \\[1.0ex]
&{}&\hspace{7mm} \times \pmatrix{e^{-i\xi} & 0 \cr 0 & e^{i\xi}}
\pmatrix{\cosh\eta & -\sinh\eta \cr -\sinh\eta & \cosh\eta } .
\end{eqnarray}

If the ray goes through $N$ cycles of this pair of layers, the S-matrix
becomes
\begin{equation}
\pmatrix{\cosh\lambda & \sinh\lambda \cr \sinh\lambda &
 \cosh\lambda }
\pmatrix{\alpha & \beta \cr \gamma & \delta}^{N}
\pmatrix{\cosh\lambda & -\sinh\lambda \cr -\sinh\lambda &
 \cosh\lambda } .
\end{equation}
Thus, the problem reduces to looking into unusual properties of
the core matrix
\begin{equation}\label{core}
\pmatrix{\alpha & \beta \cr \gamma & \delta}^{N} .
\end{equation}
We realize that the numerical computation of this expression is
rather trivial these days, but we are still interested in the
mathematical form which takes exceptionally simple form.  It is
still an interesting problem to produce mathematics which enables us
to perform calculations without using computers.  In Sec. \ref{math},
we shall consider mathematical simplification coming from one
vanishing off-diagonal element.

\section{Mathematical Instrument}\label{math}
The core matrix of Eq.(\ref{core}) contains the chain of the matrices
\begin{equation}\label{www}
W =\pmatrix{e^{-i\phi} & 0 \cr 0 & e^{i\phi}}
\pmatrix{\cosh\eta & \sinh\eta \cr \sinh\eta & \cosh\eta}
\pmatrix{e^{-i\xi} & 0 \cr 0 & e^{i\xi}}  .
\end{equation}
The Lorentz group allows us to simplify this expression under
certain conditions.

For this purpose, we transform the above expression into a more
convenient form, by taking the conjugate of each of the matrices with
\begin{equation}
C_{1} =  {1 \over \sqrt{2}} \pmatrix{1 & i \cr i & 1} .
\end{equation}
Then $C_{1} W C_{1}^{-1}$ leads to
\begin{equation}
\pmatrix{\cos\phi & -\sin\phi \cr \sin\phi & \cos\phi}
\pmatrix{\cosh\eta & \sinh\eta \cr \sinh\eta & \cosh\eta}
\pmatrix{\cos\xi & -\sin\xi \cr \sin\xi & \cos\xi} .
\end{equation}
In this way, we have converted $W$ of Eq.(\ref{www}) into a real
matrix, but it is not simple enough.

Let us take another conjugate with
\begin{equation}
C_{2} =  {1 \over \sqrt{2}} \pmatrix{1 & 1 \cr -1 & 1} .
\end{equation}
Then the conjugate $C_{2} C_{1} W C_{1}^{-1} C_{2}^{-1} $ becomes
\begin{equation}\label{abcd}
\pmatrix{\cos\phi & -\sin\phi \cr \sin\phi & \cos\phi}
\pmatrix{e^{\eta} &  0 \cr 0 & e^{-\eta}}
\pmatrix{\cos\xi & -\sin\xi \cr \sin\xi & \cos\xi} .
\end{equation}
The combined effect of $C_{2}C_{1}$ is
\begin{equation}\label{ccc}
C = C_{2}C_{1} = {1 \over \sqrt{2}} \pmatrix{e^{i\pi/4} &  e^{i\pi/4} \cr
-e^{-i\pi/4} & e^{-i\pi/4}} ,
\end{equation}
with
\begin{equation}
C^{-1} = {1 \over \sqrt{2}} \pmatrix{e^{-i\pi/4} &  -e^{i\pi/4} \cr
e^{-i\pi/4} & e^{i\pi/4}} .
\end{equation}

After multiplication, the matrix of Eq.(\ref{abcd}) will take the form
\begin{equation}
V = \pmatrix{A &  B \cr C & D} ,
\end{equation}
where $A, B, C,$ and $D$ are real numbers.  If $B$ and $C$ vanish, this
matrix will become diagonal, and the problem will become too simple.
If, on the other hand, only one of these two elements become zero, we will
achieve a substantial mathematical simplification and will be encouraged
to look for physical circumstances which will lead to this simplification.

Let us summarize.  we started in this section with the matrix
representation $W$ given in Eq.(\ref{www}).  This form can be transformed
into the $V$ matrix of Eq.(\ref{abcd}) through the conjugate
transformation
\begin{equation}\label{conju11}
V = C W C^{-1} ,
\end{equation}
where $C$ is given in Eq.(\ref{ccc}).  Conversely, we can recover
the $W$ representation by
\begin{equation}\label{conj22}
W = C^{-1} V C .
\end{equation}
For calculational purposes, the $V$ representation is much easier
because we are dealing with real numbers.  On the other hand, the
$W$ representation is of the form for the S-matrix we intend to compute.
It is gratifying to see that they are equivalent.

Let us go back to Eq.(\ref{abcd}) and consider the case where
the angles $\phi$ and $\xi$ satisfy the following constraints.
\begin{equation}\label{2angs}
\phi + \xi = 2\theta, \qquad \phi - \xi = \pi/2 ,
\end{equation}
thus
\begin{equation}\label{phixi}
\phi =  \theta + \pi/4,  \qquad   \xi = \theta - \pi/4 .
\end{equation}
Then in terms of $\theta$, we can reduce the matrix of Eq.(\ref{abcd})
to the form
\begin{equation}\label{abcd2}
\pmatrix{(\cosh\eta)\cos(2\theta)  &
       \sinh\eta - (\cosh\eta)\sin(2\theta)  \cr
       \sinh\eta + (\cosh\eta)\sin(2\theta)  &
(\cosh\eta)\cos(2\theta) } .
\end{equation}
Thus the matrix takes a surprisingly simple form if the parameters
$\theta$ and $\eta$ satisfy the constraint
\begin{equation}\label{constr}
\sinh\eta = (\cosh\eta)\sin(2\theta) .
\end{equation}\
Then the matrix becomes
\begin{equation}\label{decom2}
\pmatrix{1  &  0  \cr 2\sinh\eta & 1 }  .
\end{equation}
This aspect of the Lorentz group is known as the Iwasawa
decomposition~\cite{iwa49}, and has been discussed in the optics
literature~\cite{simon98,hkn99}.

The matrices of the form is not so strange in optics.  In para-axial
lens optics, the translation and lens matrices are written as
\begin{equation}\label{shear1}
\pmatrix{1 & u \cr 0 & 1} , \quad and  \quad \pmatrix{1 & 0 \cr u & 1} ,
\end{equation}
respectively.  These matrices have the following interesting mathematical
property~\cite{hkn97}.
\begin{equation}
\pmatrix{1 & u_{1} \cr 0 & 1} \pmatrix{1 & u_{2} \cr 0 & 1} =
\pmatrix{1 & u_{1} + u_{2} \cr 0 & 1} ,
\end{equation}
and
\begin{equation}
\pmatrix{1 & 0 \cr u_{1} & 1} \pmatrix{1 & 0 \cr u_{1} & 1}
= \pmatrix{1 & 0 \cr u_{1} + u_{2} & 1} .
\end{equation}
We note that the multiplication is commutative, and the parameter
becomes additive.  These matrices convert multiplication into
addition, as logarithmic functions do.

\section{Possible Experiments}\label{exp}

The question then is whether it is possible to construct optical layers
which will perform this or similar calculation.  In order to make
contacts with the real world, let us extend the algebra to the form
\begin{equation}
\pmatrix{1 & 0 \cr 2 \sinh\eta & 1}
\pmatrix{e^{-\eta} & 0 \cr 0 & e^{\eta}} ,
\end{equation}
which becomes
\begin{equation}
\pmatrix{e^{-\eta} & 0 \cr 2 e^{-\eta} \sinh\eta & e^{\eta} } .
\end{equation}
The square of this matrix is
\begin{equation}
\pmatrix{e^{-\eta} & 0 \cr 2 e^{-\eta} \sinh\eta  & e^{\eta}}^{2}
 = \pmatrix{e^{-2\eta} & 0 \cr 2 (e^{-2\eta} + 1)
 \sinh\eta  &  e^{2\eta} } .
\end{equation}
If we repeat this process,
\begin{equation}\label{decom3}
\pmatrix{e^{-\eta} & 0 \cr 2 e^{-\eta} \sinh\eta  & e^{\eta}}^{N}
 = \pmatrix{e^{N\eta} &  0  \cr 2b(\sinh\eta)  & e^{-N\eta} } ,
\end{equation}
with
\begin{equation}
b = e^{-N\eta} \sum_{k = 1}^{N-1}  e^{-2(k-1)\eta} ,
\end{equation}
which can be simplified to
\begin{equation}
b =  {e^{-\eta} \sinh(N\eta) \over \sinh\eta }  .
\end{equation}
\widetext
\noindent Then we can write Eq.(\ref{decom3}) as
\begin{equation}
\pmatrix{e^{-\eta} & 0 \cr 2 e^{-\eta} \sinh\eta & e^{\eta}}^{N}
 = \pmatrix{e^{-N\eta} & 0 \cr 2 e^{-\eta} \sinh(N\eta) & e^{N\eta} } .
\end{equation}
If we take into account the boundary between the air and the first medium,
\begin{equation}
\pmatrix{e^{\lambda} & 0 \cr 0 & e^{-\lambda} }
 \pmatrix{e^{-N\eta}  &  0  \cr 2 e^{-\eta} \sinh(N\eta)  & e^{N\eta} }
\pmatrix{e^{-\lambda} & 0 \cr 0 & e^{\lambda} }
= \pmatrix{e^{-N\eta}  &  0  \cr 2 e^{-(2\lambda + \eta) } \sinh(N\eta)
& e^{N\eta} }  .
\end{equation}
Thus, the original matrix of Eq.(\ref{smat}) becomes
\begin{equation}
\pmatrix{\cosh(N\eta) + i e^{-(\eta + 2\lambda) } \sinh(N\eta) &
  - (1 + i e^{-(\eta + 2\lambda) }) \sinh(N\eta) \cr
  - (1 - i e^{-(\eta + 2\lambda) }) \sinh(N\eta)
   & \cosh(N\eta) - i e^{-(\eta - 2\lambda) } \sinh(N\eta) } .
\end{equation}
\narrowtext
From the S-matrix formalism, the reflection and transmission
coefficients are
\begin{eqnarray}
&{}& R = {E^{(-)}_{a} \over E^{(+)}_{a} } =
 {S_{21} \over S_{11}}  , \nonumber \\[1ex]
&{}& T = {E^{(+)}_{s} \over E^{(+)}_{a} } =
{1 \over S_{21}}  .
\end{eqnarray}
Thus, they become
\begin{eqnarray}\label{coeff2}
&{}& R = { (1 - i e^{-(\eta + 2\lambda) }) \sinh(N\eta) \over
\cosh(N\eta) + i e^{-(\eta + 2\lambda) } \sinh(N\eta) } ,
                            \nonumber \\[1ex]
&{}& T = {- 1 \over (1 - i e^{-(\eta + 2\lambda) }) \sinh(N\eta)  }  .
\end{eqnarray}

The above expression depends only the number of layer cycles $N$ and
the parameter $\eta$, which was defined in terms of the reflection and
transmission coefficients in Eq.(\ref{coeff}).
It is important also that the above simple form is possible only if the
phase-shift parameters $\phi$ and $\xi$ should satisfy the relations
given in Eq.(\ref{phixi}) and Eq.(\ref{constr}).  In summary, they
should satisfy
\begin{equation}\label{twocon}
\cos(2\xi) = -\cos(2\phi) , \quad and \quad \tanh\eta = \cos(2\xi) .
\end{equation}
In setting up the experiment, we note that all three parameters
$\eta, \xi$ and $\eta$ depend on the incident angle and the frequency
of the light wave.  The parameter $\eta$ is derivable from the
reflection and transmission coefficients which depend on both
the angle and frequency.  The angular parameters $\xi$ and $\phi$
depend on the optical path and the index of refraction which depend
on the incident angle and the frequency respectively.

Now all three quantities in Eq.(\ref{twocon}) are functions of
the incident angle and the frequency.   If we consider a
three-dimensional space with the incident angle and frequency as
the $x$ and $y$ axes respectively.  All three quantities,
$\cos(2\xi)$, $\cos(2\phi)$, and $\tanh\eta$, will be represented
by two-dimensional surfaces.  If we choose $\cos(2\xi)$ and
$\cos(2\phi)$, the intersection will be a line.  This line will pass
through the third surface for $\tanh\eta$.  The point at which the
line passes through the  surface
corresponds to the values of the incident angle and frequency which will
satisfy the two conditions given in Eq.(\ref{twocon}).

\section*{Concluding Remarks}
In this paper, we borrowed the concept of Iwasawa decomposition from
well-known theorems in group theory.  On the other hand, group
theory appears in this paper in the form of two-by-two matrices
with three independent parameters.  The Iwasawa decomposition makes
the algebra of two-by-two matrices even simpler.  It is interesting to
note that there still is a room for mathematical simplifications in
the algebra of two-by-two matrices and that this procedure can be
tested in optics laboratories.

\end{document}